\documentclass[reprint,10pt,apl,superscriptaddress]{revtex4-1}

\usepackage[utf8]{inputenc}
\usepackage{graphicx}
\usepackage{hyperref}
\usepackage{float}
\usepackage{amsmath}
\usepackage{amssymb}
\usepackage[T1]{fontenc}
\usepackage{hyperref} 
\usepackage{isotope} 
\usepackage{upgreek}
\usepackage{sidecap}
\usepackage{wrapfig}
\usepackage{placeins}
\usepackage{subcaption}
\usepackage{color}
\usepackage[]{numprint}
\newcommand{\Onbb}{0\mbox{$\nu\beta\beta$} decay}
\newcommand{\twonbb}{2\mbox{$\nu\beta\beta$} decay}

\usepackage{booktabs,multirow} 
\usepackage{siunitx} 

\usepackage{siunitx} 
\begin{document}
\npfourdigitsep
\title{Results of a search for neutrinoless double-beta decay using the COBRA demonstrator }

\collaboration{The COBRA Collaboration}
\author{Joachim Ebert}\affiliation{Universität Hamburg, Institut für Experimentalphysik  Luruper Chaussee 149, 22761~Hamburg} 

\author{M.~Fritts}\email{Present Address: University of Minnesota, 115 Union St S.E., Minneapolis, MN 55455, USA}

\author{Daniel Gehre}\affiliation{Technische Universität Dresden, Institut für Kern- und Teilchenphysik  Zellescher Weg 19, 01069 Dresden} 
\author{Claus G{\"o}\ss{}ling}\affiliation{Technische Universität Dortmund, Lehrstuhl für Experimentelle Physik IV  Otto-Hahn-Str.~ 4, 44221 Dortmund} 
\author{Caren Hagner}\affiliation{Universität Hamburg, Institut für Experimentalphysik  Luruper Chaussee 149, 22761~Hamburg} 
\author{Nadine Heidrich}\affiliation{Universität Hamburg, Institut für Experimentalphysik  Luruper Chaussee 149, 22761~Hamburg} 
\author{Reiner Klingenberg}\affiliation{Technische Universität Dortmund, Lehrstuhl für Experimentelle Physik IV  Otto-Hahn-Str.~ 4, 44221 Dortmund} 
\author{Kevin Kr{\"o}ninger}\affiliation{Technische Universität Dortmund, Lehrstuhl für Experimentelle Physik IV  Otto-Hahn-Str.~ 4, 44221 Dortmund} 
\author{Christian Nitsch}\affiliation{Technische Universität Dortmund, Lehrstuhl für Experimentelle Physik IV  Otto-Hahn-Str.~ 4, 44221 Dortmund}
\author{Christian Oldorf}\affiliation{Universität Hamburg, Institut für Experimentalphysik  Luruper Chaussee 149, 22761~Hamburg} 
\author{Thomas Quante}\email{thomas.quante@tu-dortmund.de}\affiliation{Technische Universität Dortmund, Lehrstuhl für Experimentelle Physik IV  Otto-Hahn-Str.~ 4, 44221 Dortmund}
\author{Silke Rajek}\affiliation{Technische Universität Dortmund, Lehrstuhl für Experimentelle Physik IV  Otto-Hahn-Str.~ 4, 44221 Dortmund} 
\author{Henning Rebber}\affiliation{Universität Hamburg, Institut für Experimentalphysik  Luruper Chaussee 149, 22761~Hamburg} 
\author{Katja Rohatsch}\affiliation{Technische Universität Dresden, Institut für Kern- und Teilchenphysik  Zellescher Weg 19, 01069 Dresden}
\author{Jan Tebr{\"u}gge}\affiliation{Technische Universität Dortmund, Lehrstuhl für Experimentelle Physik IV  Otto-Hahn-Str.~ 4, 44221 Dortmund} 
\author{Robert Temminghoff}\affiliation{Technische Universität Dortmund, Lehrstuhl für Experimentelle Physik IV  Otto-Hahn-Str.~ 4, 44221 Dortmund}
\author{Robert Theinert}\affiliation{Technische Universität Dortmund, Lehrstuhl für Experimentelle Physik IV  Otto-Hahn-Str.~ 4, 44221 Dortmund}
\author{Jan Timm}\affiliation{Universität Hamburg, Institut für Experimentalphysik  Luruper Chaussee 149, 22761~Hamburg} 
\author{Bj{\"o}rn Wonsak}\affiliation{Universität Hamburg, Institut für Experimentalphysik  Luruper Chaussee 149, 22761~Hamburg} 
\author{Stefan Zatschler}\affiliation{Technische Universität Dresden, Institut für Kern- und Teilchenphysik  Zellescher Weg 19, 01069 Dresden}
\author{Kai Zuber} \affiliation{Technische Universität Dresden, Institut für Kern- und Teilchenphysik  Zellescher Weg 19, 01069 Dresden}


\date{\today}

\pdfinfo{%
  /Title    ()
  /Author   ()
  /Creator  ()
  /Producer ()
  /Subject  ()
  /Keywords ()
}
\begin{abstract}
Neutrinoless double-$\beta$ decay (\Onbb) is a hypothetical process that can occur if the neutrino is its own antiparticle. The COBRA collaboration operates a demonstrator to search for these decays at the Laboratori Nazionali del Gran Sasso in Italy using CdZnTe semiconductor detectors. The exposure of $234.7\,$kg\,d considered in this analysis was collected between September 2011 and February 2015. The analysis focuses on the decay of the nuclides \isotope[114][]{Cd}, \isotope[128][]{Te}, \isotope[70][]{Zn}, \isotope[130][]{Te} and \isotope[116][]{Cd}.  A Bayesian analysis is performed to estimate the signal strength of \Onbb. No signal is observed for any of these nuclides. Therefore, the following half-life limits at 90\,\% credibility are set: $T_{1/2}^{0\nu}>1.6\times10^{21}\,$yr (\isotope[114][]{Cd}), $T_{1/2}^{0\nu}>1.9\times10^{21}\,$yr (\isotope[128][]{Te}), $T_{1/2}^{0\nu}>6.8\times10^{18}\,$yr (\isotope[70][]{Zn}), $T_{1/2}^{0\nu}>6.1\times10^{21}\,$yr (\isotope[130][]{Te}), and $T_{1/2}^{0\nu}>1.1\times10^{21}\,$yr (\isotope[116][]{Cd}).
\end{abstract}

\maketitle


\section{Introduction}
An open question in neutrino physics is the absolute mass and nature of the neutrino. Is it a Majorana or a Dirac particle, i.e. is it its own anti-particle or not?  These question can potentially be answered if neutrinoless double-$\beta$ decay (\Onbb) is observed.

Double-$\beta$ decay is a rare second-order weak process that changes the atomic number by 2 units, while leaving the atomic mass constant \cite{PhysRev.48.512}. It comes in two modes: neutrino-accompanied double-$\beta$ decay (\twonbb) results in two charged leptons and the corresponding neutrinos, $(A,Z)\rightarrow(A,Z+2)+2e^{-} +2\bar{\nu}$. This decay mode has been observed in several isotopes, and half-lives between $7.1\times 10^{18}$\,yr (\isotope[100][]{Mo}) and $7.7\times 10^{24}$\,yr (\isotope[128][]{Te}) have been measured \cite{Barabash:2010bd,PhysRevC.81.035501,PhysRevC.78.054606}. Neutrinoless double-$\beta$ decay \cite{PhysRev.56.1184}, which is mediated via an exchange of a virtual neutrino, can only occur if the neutrino is a Majorana particle \cite{Majorana:1937vz}. This process is forbidden in the standard model of particle physics because it violates the lepton number by 2 units: $(A,Z)\rightarrow(A,Z\pm2)+2e^{\mp}$. The half-life of this decay is inversely proportional to the square of the 
effective  Majorana neutrino mass. 

\Onbb\, has been searched for in several isotopes, but has not been observed, yet. Recent experiments report upper limits on 
the half-life. The GERDA collaboration searches for \Onbb\, of \isotope[76][]{Ge} using high purity germanium detectors; the currently best limit on the half-life is $T^{0\nu}_{1/2}>1.9\times10^{25}\,$yr at 90\,\% credibility \cite{Agostini:2013mzu}.
The KamLAND-Zen and EXO-200 collaborations have searched for the \Onbb\, of \isotope[136][]{Xe} \cite{kamland_zen_2014,exo_200_results_2014}. Both experiments have observed zero events in their region of interest, which results in a half-life lower limit of $T^{0\nu}_{1/2}>2.6\times10^{25}\,$\,yr  for the KamLAND-Zen Collaboration and  $T^{0\nu}_{1/2}>1.1\times10^{25}\,$yr for the EXO-200 Collaboration, both at a 90\,\% confidence level. The CUORE collaboration set a new limit of $T^{0\nu}_{1/2}>4\times10^{24}$\,yr at 90\,\% credibility for  \Onbb\,of \isotope[130][]{Te} \cite{Alfonso:2015wka}. The Solotvina experiment sets a limit of $T^{0\nu}_{1/2}>1.7\times10^{23}\,$yr for \isotope[116][]{Cd} \cite{PhysRevC.68.035501}. Other experiments have searched for the \Onbb\, of \isotope[114][]{Cd} and \isotope[70][]{Zn} by using CdWO$_{2}$ and ZnWO$_{2}$ scintillator crystals. Both experiments have observed zero events which results in half-life limits of $T^{0\nu}_{1/2}>1.1\times10^{21}\,$yr for \isotope[114][]{Cd} and $T^{0\nu}_{1/2}>3.2\times10^{19}\,$yr for \isotope[70][]{Zn}, both limits are reported at 90\,\% confidence level \cite{cdwo,0954-3899-38-11-115107}. For an overview of the current experimental status, see e.g. Ref. \cite{Maneschg2015188}.
\newline
\newline
This paper reports on the search for \Onbb\, of several isotopes using the COBRA demonstrator. The CdZnTe semiconductor crystals deployed in the setup contain nine double-$\beta$ isotopes with several decay modes. A focus is placed on the five $\beta^-\beta^-$ ground-state-to-ground-state transitions of the isotopes \isotope[114][]{Cd}, \isotope[128][]{Te}, \isotope[70][]{Zn},  \isotope[130][]{Te} and \isotope[116][]{Cd} with $Q$-values between 542\,keV and \numprint{2813}\,keV. Section\,II introduces the COBRA demonstrator. The data taking and data quality criteria are discussed in Section\,III. In Section\,IV,  the energy reconstruction and the calibration procedure are  explained. The pulse shape analysis is discussed in Section\,V. The event selection for the analysis is discussed in Section\,VI. Section\,VII explains the methods used for the signal estimation. Section\,VIII introduces all systematic uncertainties which are considered in the signal estimation. The results are discussed in Section\,IX and 
Section\,X concludes the article.

\section{The COBRA Demonstrator}

The COBRA demonstrator is located at the Laboratori Nazionali del Gran Sasso (LNGS) of the INFN in L'Aquila, Italy. The setup is described in detail in Ref. \cite{Ebert:2015uta}. The demonstrator comprises 64 CdZnTe coplanar-grid semiconductor detectors arranged in four layers of $4\times4$ detectors. Each detector has a size of $1\times1\times1\,$cm$^3$ and a mass of $5.9\,$g. The detectors are supported by a polyoxymethylene frame installed in a support structure made of electroformed copper. The setup is constantly flushed with evaporated nitrogen to suppress radon-induced background. The detectors are surrounded by 5\,cm of electroformed copper, followed by a 5-cm layer of ultra-low activity lead (<3\,Bq/\,kg of \isotope[210][]{Pb}) and 15\,cm of standard lead. A $5$-cm borated polyethylene layer is used to reduce the neutron flux inside the setup. A charge-sensitive preamplifier integrates the current pulses and converts the single-ended detector pulses into differential signals in order to suppress 
electronic noise during transmission. After linear amplification, the pulse shapes are digitized using $100\,$MHz flash analog-to-digital converters (FADCs) and written to disk.

\section{Data Taking and Data Quality criteria}

The data taking of the COBRA demonstrator started in September 2011 after the commissioning of a partial setup that contained only one detector layer. The other three layers were installed successively in 2012 and 2013. The demonstrator was completed in November 2013 with 64 installed detectors. Currently, 61 out of 64 detectors operate under stable conditions. For further information see Ref. \cite{Ebert:2015aia}.

This publication uses data corresponding to an exposure of $234.7\,$kg\,d, recorded between September 2011 and February 2015. A set of quality criteria based on the observed noise level of single detectors is applied. A small fraction of data taking runs is discarded because of a failure of the nitrogen flushing. In such cases radon diffused into the setup and the overall background rate increased for the duration of these runs. After the data-quality selection, the dataset corresponds to an exposure of 216.1\,kg\,d.

For the analysis of \isotope[114][]{Cd}, runs with energy thresholds larger than $421\,$keV are discarded due to the lowered trigger efficiencies. Therefore, the corresponding exposure is  reduced to $212.8$\,kg\,d.

\section{Energy Calibration and Resolution}

The energy deposited inside a CdZnTe detector is estimated by the pulse heights of signals measured with electrodes on the detector surface. Because the mobility-lifetime product for holes in CdZnTe is 2 orders of magnitude lower than for electrons, the hole contribution to the induced signal depends strongly on the interaction depth. Therefore, CdZnTe detectors are typically designed to have a coplanar-grid configuration, which is a modification of the Frisch grid principle used for semiconductors \cite{cpg94}. This approach is based on two interleaved comb-shaped anodes, a collecting anode (CA) on ground potential and a non-collecting anode (NCA) on a slightly negative potential. The deposited energy is then estimated to be
\begin{equation}
    E\propto H_{\text{CA}}-\omega H_{\text{NCA}},
\end{equation}
where $H_{\text{CA}}$ and $H_{\text{NCA}}$ are the pulse heights of the two anode signals and $\omega$ is a weighting factor introduced to compensate for electron trapping. This weighting factor is determined and optimized for each detector by calibration measurements. In this configuration, the signal shape is only generated by the electron contribution.  Besides the improved energy resolution, the coplanar-grid technique allows for the reconstruction of the interaction depth. An analytical model for the calculation of the interaction depth is discussed, e.g., in Ref. \cite{FrittsCPG}.

\begin{figure}[h]
    \includegraphics[width=0.5\textwidth]{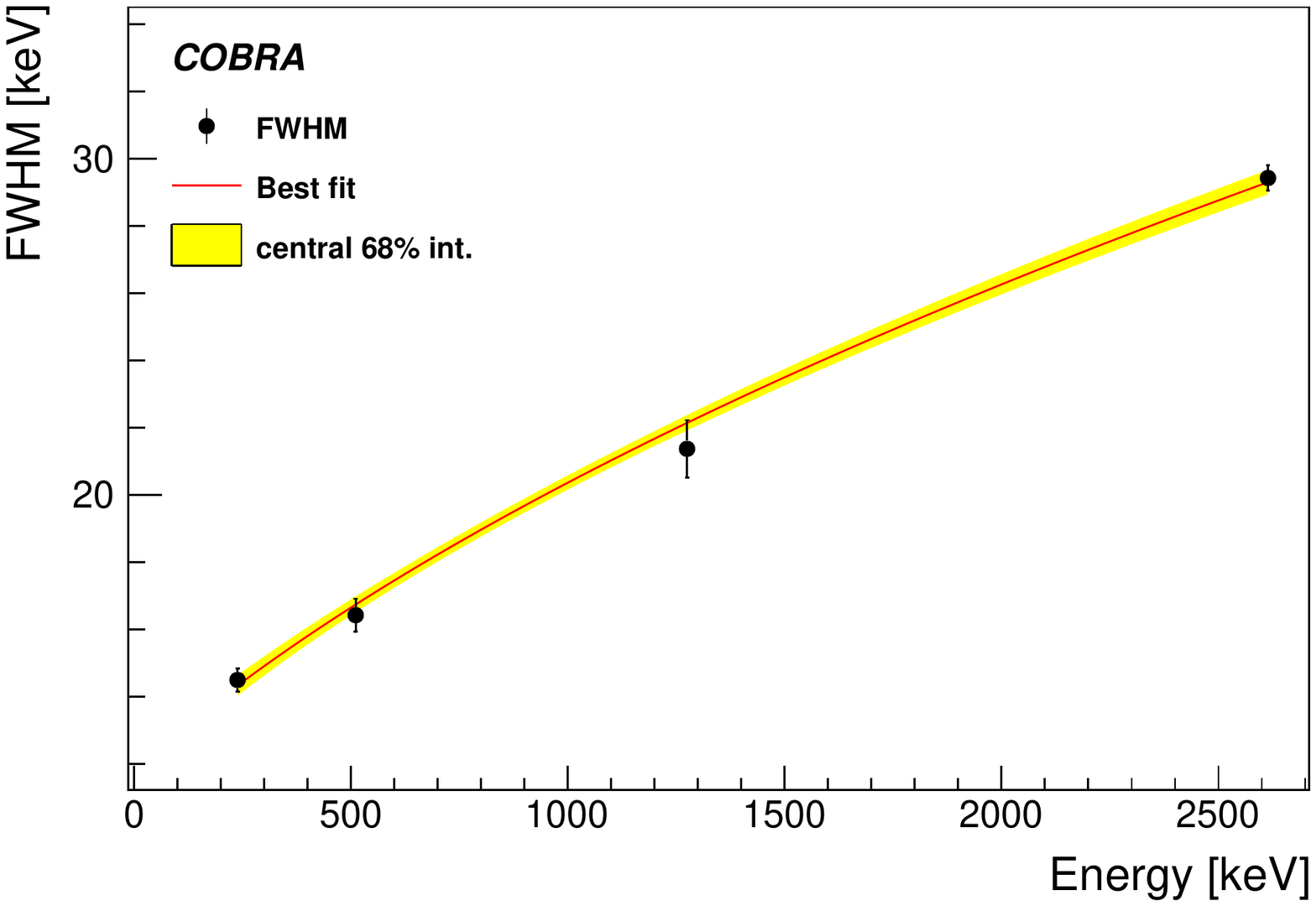}
    \caption{The effective FWHM as a function of the energy determined with calibration measurements (markers). The used line energies are $238\,$\,keV (\isotope[212][]{Pb}), $511$\,keV (annihilation $\gamma$ radiation), $1\,275$\,keV (\isotope[22][]{Na}) and $2\,614$\,keV (\isotope[208][]{Tl}). Also shown is the parametrization (red line) including the corresponding uncertainty (yellow band). }
    \label{fig:rescal_fit}
\end{figure}

A \isotope[22][]{Na} and a \isotope[228][]{Th} source are used to determine the energy calibration of each detector. The energy threshold of each detector is set depending on the noise level. The thresholds range from $40\,$keV  up to $550$\,keV for all ROIs except the \isotope[114][]{Cd} ROI, where the highest threshold is 421\,keV. Higher noise levels are observed after periods of detector maintenance if the detectors have been exposed to light or humidity.

The effective energy resolution of the demonstrator setup is estimated from a combined spectrum of all detectors, each weighted with its corresponding exposure. The $\gamma$ peaks from measurements with the calibration sources are modeled by double-Gaussian functions from which the full width at half maximum (FWHM) is calculated. The parametrization of the energy resolution as a function of the energy takes into account the charge carrier production and a noise component. The FWHM takes the form
\begin{equation}
    \text{FWHM}(E)=\sqrt{aE + b},
\end{equation}
where $a$ is a Poisson component and $b$ is associated with the noise. \autoref{fig:rescal_fit} shows the effective FWHM as a function of the incident energy. It ranges from about $15$\,keV at an energy of $238$\,keV up to approximately $30$\,keV at $2.6$\,MeV.

\begin{figure}
    \includegraphics[width=0.5\textwidth]{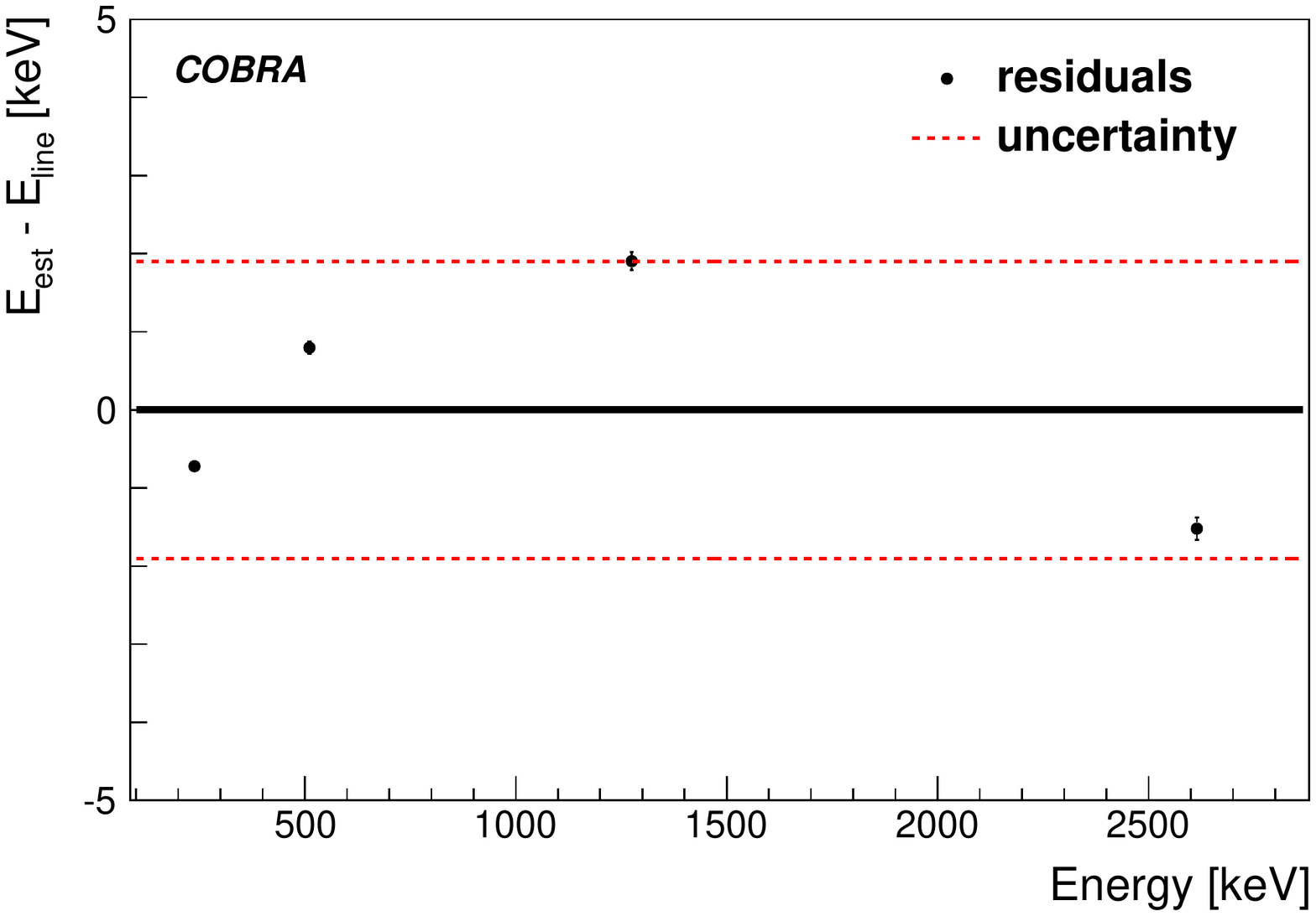}
    \caption{Difference between the estimated and the true energy for several $\gamma$ lines (markers) as well as the systematic uncertainty associated with the energy calibration (red dashed line)}
    \label{fig:cal_unc_graph}
\end{figure}

A linear function is used to determine the energy calibration of each detector. \autoref{fig:cal_unc_graph} shows the difference between the peak position and the calculated calibration function of several $\gamma$ lines.

\section{Pulse Shape Analysis}
The analysis of the pulse shapes is a powerful tool to identify energy deposits on the detector surface caused, e.g., by $\alpha$ and $\beta$ radiation. Three characteristic quantities are defined in the following, which are used to suppress events which deposit energies close to the electrodes(depth requirement) and lateral surface events (LSE requirement). The pulse shape analysis is performed using MAnTiCORE \cite{schulz}. The recorded pulse shapes can be divided into four time intervals: The prepulse baseline, the charge-cloud drift, the charge collection, and the postpulse baseline. The prepulse and post-pulse baselines of the difference pulse are used to determine the pulse height of the signal, which is proportional to the deposited energy. During the charge-cloud drift the induced signal is approximately the same on both anodes. The difference signal is therefore approximately zero. Once the charge cloud approaches the anode side the grid bias potential becomes important. The electrons are 
collected by the CA, leading to a sharp rise of the CA signal and a sharp fall of the NCA signal. The resulting weighted difference signal is a function of the energy and does not depend on the interaction depth. The interaction depth is reconstructed by the ratio of the cathode pulse height and the difference pulse height. The cathode signal itself is reconstructed from the sum of both anode signals.

If the energy is deposited on the lateral surfaces near the CA, the difference signal shows an earlier rise than the bulk signal. This behavior is characterized by the early rise time (ERT) value, which is defined as the time difference between the $3\,$\% point to $50\,$\% point of the maximum pulse height. On the NCA near lateral surfaces, the difference pulse shows a small dip below the baseline before the sharp rise of the signal. It is characterized by the dip value, which is defined as the difference of the baseline and the pulse minimum in a window of 30 samples ending at the $50\,$\% pulse height point on the right edge. More information about these quantities are given in Ref. \cite{fritts_tebr}.

\section{Event Selection}
Pulse shapes are recorded if at least one detector records an energy deposition above its energy threshold. Before any software requirement is applied, the spectrum between $\numprint{400}$\,keV and $5$\,MeV contains $\numprint{166668}$\,counts; $\numprint{104013}$\,counts remain after requiring the reconstructed interaction depth to be in the interval $[0.2,0.97]$ and thus removing events close to the cathode or the anode side. Requiring the dip and ERT value below $53$ and $8$, respectively, reduces the number of events to $\numprint{65176}$. For the current analysis, no anticoincidence between the detectors is required. This will be studied in the future. 

\begin{figure}[h]
    \includegraphics[width=0.48\textwidth]{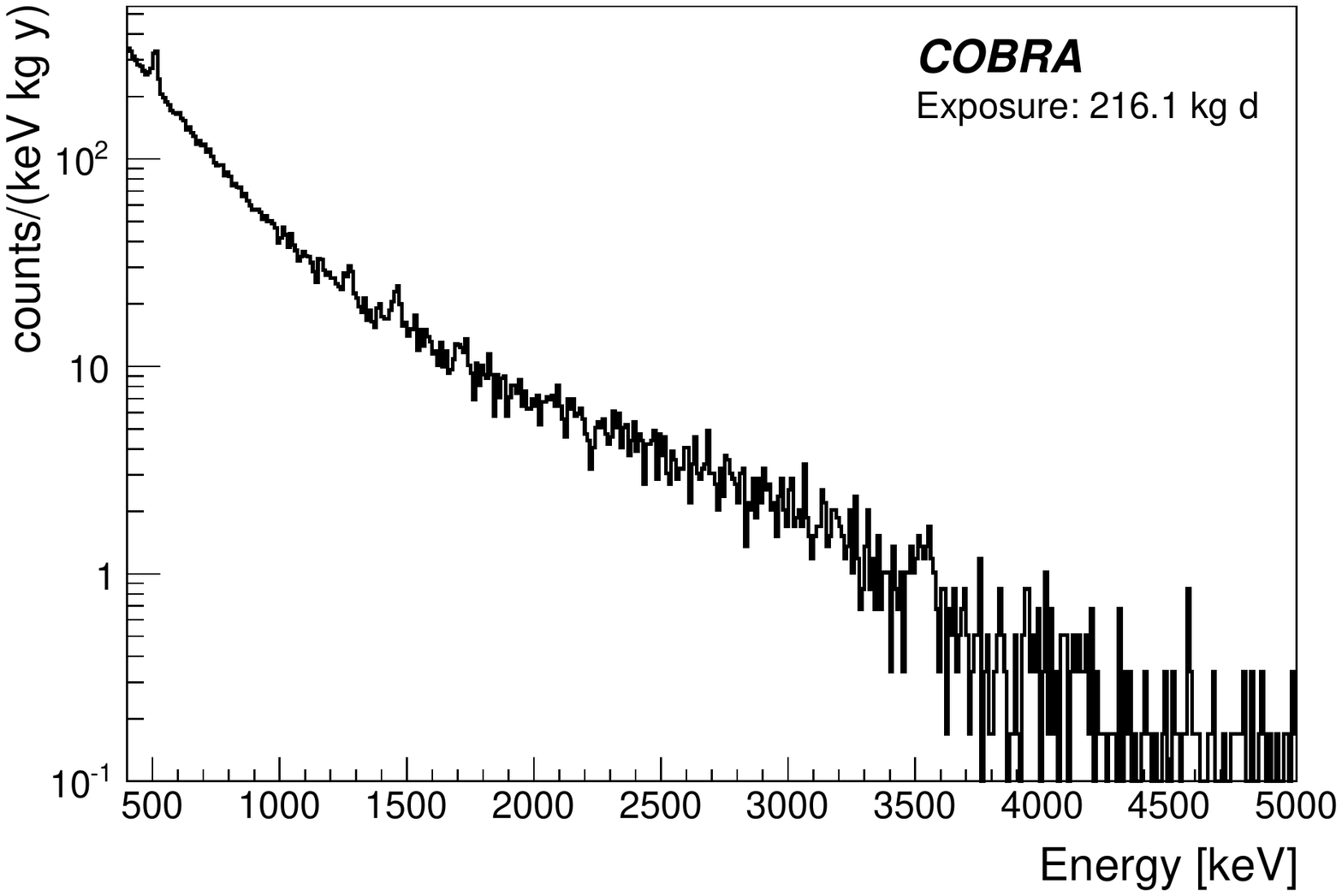}
    \caption{The energy spectrum of all detectors after the full event selection. Three $\gamma$ lines are identified: the annihilation line at $511$-keV, the $\numprint{1275}$-keV line of \isotope[22][]{Na} and the  1460-keV line of \isotope[40][]{K}. The line at $3.5$\,MeV is most likely caused by $\alpha$ radiation on the cathode side.}
    \label{fig:scaled_spectrum}
\end{figure}

\autoref{fig:scaled_spectrum} shows the spectrum between $400$\,keV and $5$\,MeV after the event selection. Three $\gamma\,$lines are identified. The $e^{+}e^{-}$ annihilation line at $\numprint{511}$\,keV is most likely dominated by \isotope[22][]{Na}, which is also responsible for the line visible at $\numprint{1275}\,$keV. A primordial source of radiation is \isotope[40][]{K} causing a line at $\numprint{1460}$\,keV. The small peak at $3.5$\,MeV is most likely caused by $\alpha$-radiation on the cathode side. They are induced by Radon remnants which are attracted to the cathode after short failures of the nitrogen flushing. The detectors of the fourth layer have a coated cathode side with a thickness of $30$\,$\mu$m. This is not sufficient to stop all $\alpha$ particles, but instead reduces their kinetic energy.

\section{Background Modeling and Signal Estimation}

A Bayesian analysis \cite{BAT} is performed to estimate the signal strength of potential \Onbb. For each isotope a ROI is defined as as the $\pm3$\,FWHM interval around the $Q$-value. For \isotope[114][]{Cd} and $\isotope[130][]{Te}$ the ROI is extended to cover the $\gamma$ lines which can contribute to the background. The statistical model is a binned likelihood in the energy distribution describing Poisson fluctuations of the number of events $n_{i}$ in each bin, i.e.
\begin{equation}
    p(\vec{n}|\vec{\nu}_{s},\vec{\nu}_{b},\vec{\nu}_{\text{line}})=\prod_{i=1}^{n}\frac{e^{-(\nu_{s}^{i}+\nu_{b}^{i}+\nu_{\text{line}}^{i})}(\nu_{s}^{i}+\nu_{b}^{i}+\nu_{\text{line}}^{i})^{n_{i}}}{n_{i}!},
\end{equation}
with $\nu_{s}^{i}$ being the expectation of the signal, $\nu_{b}^{i}$ the expectation of the continuous background in each bin $\Delta E_{i}$ and $\nu_{\text{line}}^{i}$ the expectation of a line-shaped background, which is only greater than zero if a line is present in the ROI.
 
Based on previous COBRA data, the background is modeled by an exponential function: 
\begin{equation}
    \nu_{b}^{i}=\int_{\Delta E_{i}}\frac{B\lambda}{\left( e^{-\lambda E_{\text{min}}}-e^{-\lambda E_{\text{max}}}\right)}e^{-\lambda E}dE,
\end{equation}
where $B$ is the total number of background events in the ROI, $E_{\text{min/max}}$ are the interval boundaries of the ROI and $\lambda$ is the decay parameter of the exponential function. The comparison with a background model using a second order polynomial results in a difference in the estimated half-life of less than 2\,\% 

Lines from $\gamma$ radiation and the signal process have intrinsic widths smaller than the energy resolution. The energy resolution varies from detector to detector. Thus, a $\gamma$\,line in the dataset combining all detectors should in principle be described by a superposition of 64 Gaussian peaks. For simplicity such lines are described by double-Gaussian functions which model the energy resolution of the full detector array:
\begin{equation}
    \nu_{\text{line/s}}^{i}=\int_{\Delta E_{i}}A\left(\frac{F}{{\sqrt{2\pi}\sigma_{1}}}e^{\frac{-(E-E_{0})^2}{2\sigma_{1}^2}}+\frac{1-F}{{\sqrt{2\pi}\sigma_{2}}}e^{\frac{-(E-E_{0})^2}{2\sigma_{2}^2}}\right)dE,
\end{equation}
where $\nu_{\text{line/s}}^{i}$ is the expected number of counts for an energy interval $\Delta E_{i}$, $A$ is the amplitude, $F$ is the fraction of the first Gaussian, $E_{0}$ is the peak position and $\sigma_{1,2}$ are the widths of the Gaussian functions. $A$ can be the total number of signal events, $S$, or $B_{\text{line}}$ the total number of line-shaped background events. Studies show that $\sigma_{1}=2\sigma_{2}$ is a reasonable choice which describes the full energy peaks well and ensures that the fraction of both Gaussian functions is stable over the whole energy range. 

The parameters $B$, $\lambda$, $S$ and, if applicable, $B_{\text{line}}$ are free parameters of the model, for which uniform prior probabilities are assumed. The parameters $F$, $\sigma_{1,2}$ and $E_{0}$ are nuisance parameters, which are used to model the corresponding systematic uncertainties, assuming Gaussian prior probabilities.

The half-life of \Onbb\,is calculated from the signal strength as
\begin{eqnarray}
    T_{1/2}^{0\nu\beta\beta}&=&\frac{\ln(2) \, N \, X \, \epsilon}{\hat{s}},
\end{eqnarray}
with $N$ being the number of atoms of the evaluated isotope per kilogram detector mass, $X$ is the total exposure, and $\hat{s}$ is the observed signal strength of the \Onbb. The total efficiency $\epsilon$ includes the intrinsic efficiency for the signal process and the efficiencies of the pulse shape analysis requirements. The systematic uncertainties associated with the efficiency $\epsilon$ and the number of atoms $N$ are also modeled using nuisance parameters and assuming Gaussian prior probabilities.

\section{Systematic Uncertainties}
The systematic uncertainties discussed in the following are used as the standard deviations of the Gaussian priors for the corresponding nuisance parameters. These are summarized in \autoref{tab:systematic}.

The uncertainties of the $Q$-values range from $0.01\,$keV to $2.1\,$keV \cite{qval-2012}. The systematic uncertainty associated with the energy calibration, and thus on the peak position $E_{0}$, is estimated by the largest difference between the calibration function and the means of Gaussian fits of all $\gamma$ lines. The maximum difference is $1.9\,$keV.

The total uncertainty associated with the mean of the \Onbb\,line is calculated using uncertainty propagation of $Q$-value uncertainty and the uncertainty of the energy calibration.

The uncertainties of $F$ and $\sigma_1$ are estimated via calibration measurements. The fraction $F=0.52\pm0.03$ is constant over the considered energy range. The $\sigma_{1}$ for each ROI is calculated from the resolution function shown in \autoref{fig:rescal_fit}. The values range from $7.2\,$keV up to $12.8$\,keV, and the corresponding uncertainties range from $0.09$\,keV to $0.15$\,keV.

All efficiencies are estimated as the ratio of events which fulfill the corresponding selection criteria and the total number of events. The depth requirement efficiency is evaluated with a pulse shape simulation. It is based on a simulation of the electric fields for the coplanar electrode layout which simulates charge propagation including diffusion and electromagnetic repulsion of the charge cloud. To estimate possible systematic deviations from the data, the results of the simulation are compared to an efficiency calculated using the observed \isotope[113][]{Cd} spectrum with an endpoint of $320$\,keV. For further information about the spectrum see Ref. \cite{Ebert:2015aia}. \isotope[113][]{Cd} is homogenously distributed inside the detector and gives a good measure of the efficiency losses caused by the depth requirement in an energy range from $\numprint{180}\,$keV to $\numprint{300}\,$keV. The absolute difference between both efficiency measurements is about $0.1$. A possible source of these systematic deviations are distortions of the energy reconstruction near the anode grids. The efficiency is calculated for each ROI separately and 
range from $0.74$ (\isotope[114][]{Cd}) to $0.78$ (\isotope[116][]{Cd}). The associated absolute systematic uncertainty is $\pm0.1$. 

The impact of the LSE requirement on the efficiency is estimated with the double escape peak of the \isotope[208][]{Tl} $\gamma$ line of the calibration measurements. This is advantageous because the early rise time also depends on the spatial distribution of the energy deposits. This distribution is smaller for the double escape peak and \Onbb\, events and it is more confined than events caused by multiple-scattered photons. The efficiency is systematically underestimated even if calculated at the double escape peak because of the underlying background caused by \isotope[212][]{Bi}. The emitted gamma line overlaps strongly with the double escape peak and hence the fraction of the \isotope[212][]{Bi} line is minimized by using the low energy side of the peak. Based on measurements with calibration data the efficiency is estimated to be $0.76\pm0.02 $.

The intrinsic efficiency is the probability to deposit the whole energy of a $0\nu\beta\beta$ event inside the active volume of one detector and is estimated with Monte-Carlo simulations. The values range from $0.62$ to $0.96$. The absolute uncertainty of the calculated efficiencies is not expected to be larger than $0.002$. The simulation is based on GEANT4 9.6, which models the particle propagation and interaction with matter \cite{geant}. The energy and angular distributions of the physics processes are predicted by an event generator based on DECAY0 \cite{decay0}. To estimate detector specific efficiency losses, causes, e.g., by dead layers, each detector is scanned with a \isotope[137][]{Cs} point source with a known activity in three different distances. The peak count rates of these three measurements are fitted with $f(r)=a/r^{2}$. The fit parameter $a$ is proportional to the activity of the source, the solid angle fraction covered by the detector, the emission probability of the $\gamma$-line and 
the efficiency of the detector. For the estimation of the efficiency of the full detector array, the detector with the highest efficiency is assumed to be fully efficient. All other detector efficiencies are calculated in relation to this detector. The average detection efficiency of the detector array is $0.89\pm0.01$.  

\begin{table}[htb]
    \caption{The first part lists all means and widths which are used as prior informations. The second part lists all values which are used to calculate the total uncertainties.} 
    \label{tab:systematic} 
    \centering 
    \begin{tabular}{
    l
    c
    S[table-format=1.3,table-comparator=true,table-space-text-post={}] 
    S[table-format=1.3,table-comparator=true,table-space-text-post={}] 
    S[table-format=1.3,table-comparator=true,table-space-text-post={}] 
    S[table-format=1.3,table-comparator=true,table-space-text-post={}] 
    S[table-format=1.3,table-comparator=true,table-space-text-post={}] 
    }\toprule\toprule
         & &\multicolumn{0}{c}{\isotope[114][]{Cd}}  & \multicolumn{0}{c}{\isotope[128][]{Te}} & \multicolumn{0}{c}{\isotope[70][]{Zn}} & \multicolumn{0}{c}{\isotope[130][]{Te}} & \multicolumn{0}{c}{\isotope[116][]{Cd}} \\\toprule
        $Q$-value&[keV] & 543 & 867 & 997 & 2528 &2814\\
        $\sigma_{Q\text{-tot}}$&[keV] &2.1 &2.1 &2.8 &1.9 &2 \\
        $F$ & & 0.52 & 0.52 & 0.52 & 0.52 & 0.52\\
        $\sigma_{F}$ & & 0.03 &  0.03 & 0.03  & 0.03 &0.03\\
        $\sigma_1$ & [keV] &7.23 &8.27 &8.65 &12.29 &12.86 \\
        $\sigma_{1,\text{unc}}$ &[keV]&0.1 &0.09 &0.09 &0.15 & 0.15\\
        $\epsilon_{\text{tot}}$ & & 0.48 & 0.46 & 0.45 & 0.34 & 0.33 \\
        $\sigma_{\epsilon_{\text{tot}}}$& & 0.07 & 0.07  & 0.07 & 0.05 &0.05\\
        $N$ &$[10^{23}\frac{\text{atoms}}{\text{kg}}]$ & 6.59 & 8.08 & 0.015 & 8.62 & 1.73\\
        $\sigma_{N}$ &$[10^{23}\frac{\text{atoms}}{\text{kg}}]$& 0.132 & 0  & 0.003 & 0& 0.035\\
        & & & & & &\\
        $\sigma_{Q\text{-value}}$& [keV]& 0.9 & 0.9 & 2.1 & 0.01 &  0.13 \\
        $\sigma_{cal}$&[keV] &1.9 &1.9 &1.9 &1.9 &1.9\\
        $\epsilon_{\text{depth}} $ & & 0.74& 0.74&0.74 &0.76 &0.78\\
        $\sigma_{\epsilon_{\text{depth}}}$ & &0.1 & 0.1&0.1 &0.1 &0.1\\
        $ \epsilon_{lse}$ &  & 0.76 & 0.76 & 0.76 & 0.76&0.76\\
        $\sigma_{\epsilon_{lse}}$& & 0.02 & 0.02 & 0.02 & 0.02 &0.02\\
        $\epsilon_{int} $ & &0.96 &0.92 &0.90 &0.66 &0.62\\
        $\sigma_{\epsilon_{int}}$& & 0.002 & 0.002 & 0.002 &  0.002&0.002\\
        $\epsilon_{det} $ & &0.89 &0.89 &0.89 &0.89 &0.89\\
        $\sigma_{\epsilon_{det}} $ & &0.01 &0.01 &0.01 &0.01 & 0.01\\\bottomrule\bottomrule
        \end{tabular}
\end{table}

The total efficiency uncertainty is obtained by error propagation and is shown in \autoref{tab:results}. Runs with the highest trigger threshold of $550\,$keV are excluded for the \isotope[114][]{Cd} analysis and are well below all other region of interests. Therefore the trigger efficiency is assumed to be 100\%.

One virtual CdZnTe molecule consists of $\text{Cd}_{0.9}\text{Zn}_{0.1}\text{Te}_{1}$. This composition is consistent  with inductively coupled plasma mass spectroscopy measurements, which were done with a CdZnTe sample. However, the fraction of Zn can vary throughout the crystals and therefore a systematic uncertainty of $\pm2\,$\% points is introduced for the Cd fraction, which corresponds to a variation of $\pm20\,\% $ in the Zn fraction.

The detector lifetime is measured by the internal FADC clock. The uncertainty of each clock is of the order of $10\frac{\text{ms}}{4\text{h}}$.  The mass of each detector is measured before installation with an
accuracy of milligrams which is negligible compared to a single detector mass of approx $6$\,g. Therefore, the exposure uncertainty is neglected in this analysis.

\section{Results}
\begin{figure*}
    \begin{subfigure}[]{0.49\textwidth}
        \includegraphics[width=\textwidth]{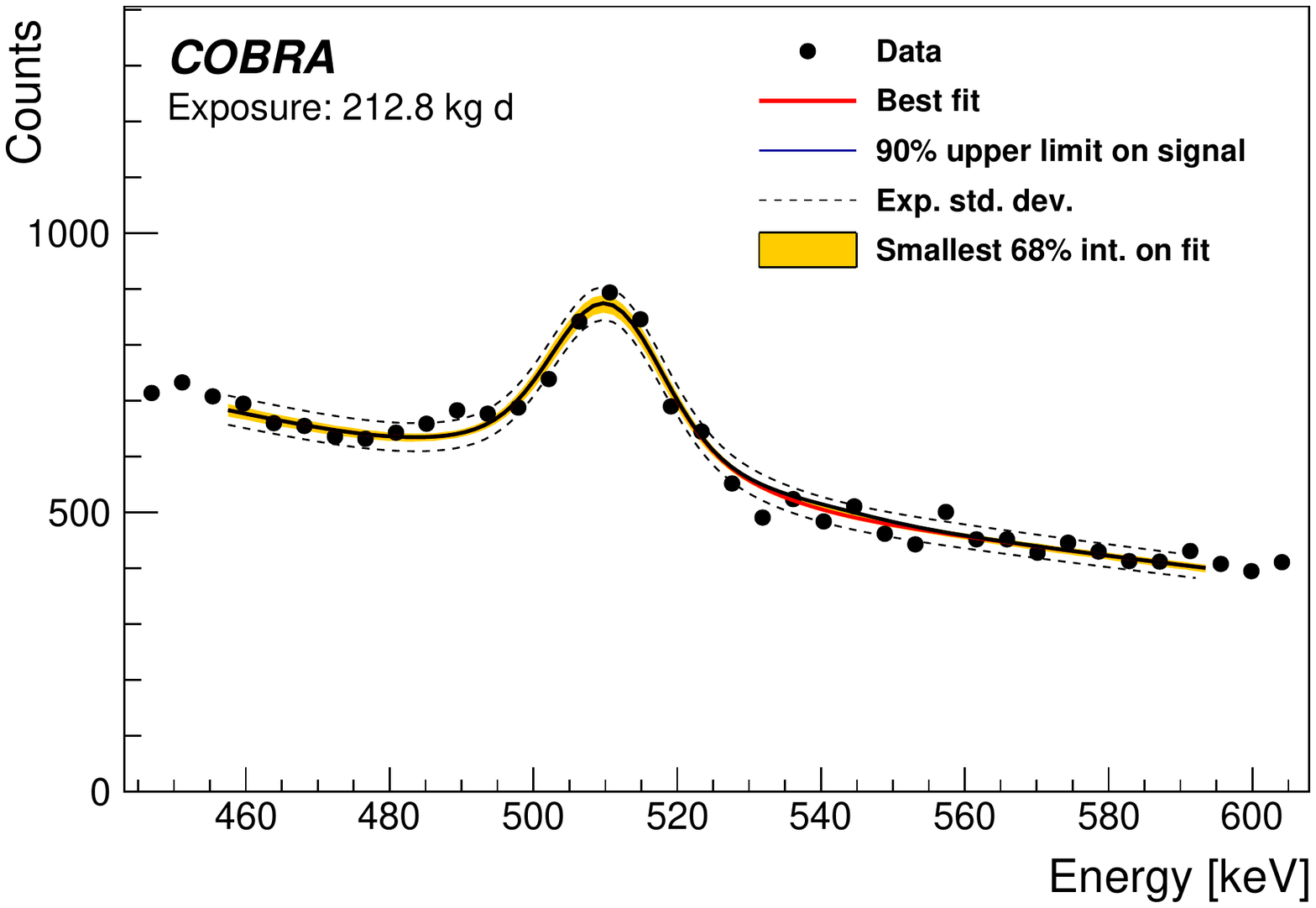}
        \caption{\isotope[114][]{Cd}, including the 511\, keV line} 
        \label{fig:cd114fit}
   \end{subfigure}
   \begin{subfigure}[]{0.49\textwidth}
        \includegraphics[width=\textwidth]{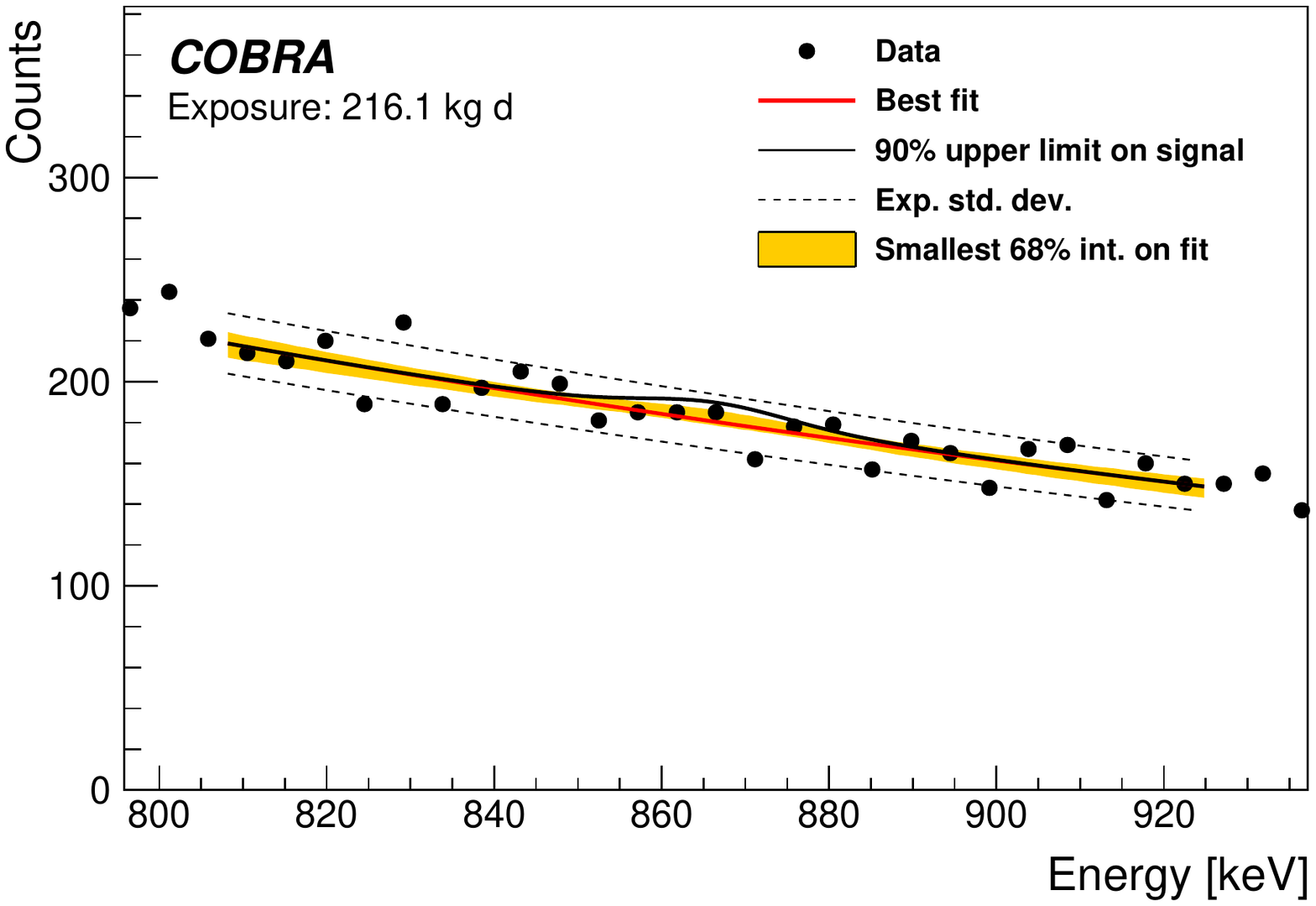}
        \caption{\isotope[128][]{Te}}
        \label{fig:te128fit}
   \end{subfigure}
   \begin{subfigure}[]{0.49\textwidth}
        \includegraphics[width=\textwidth]{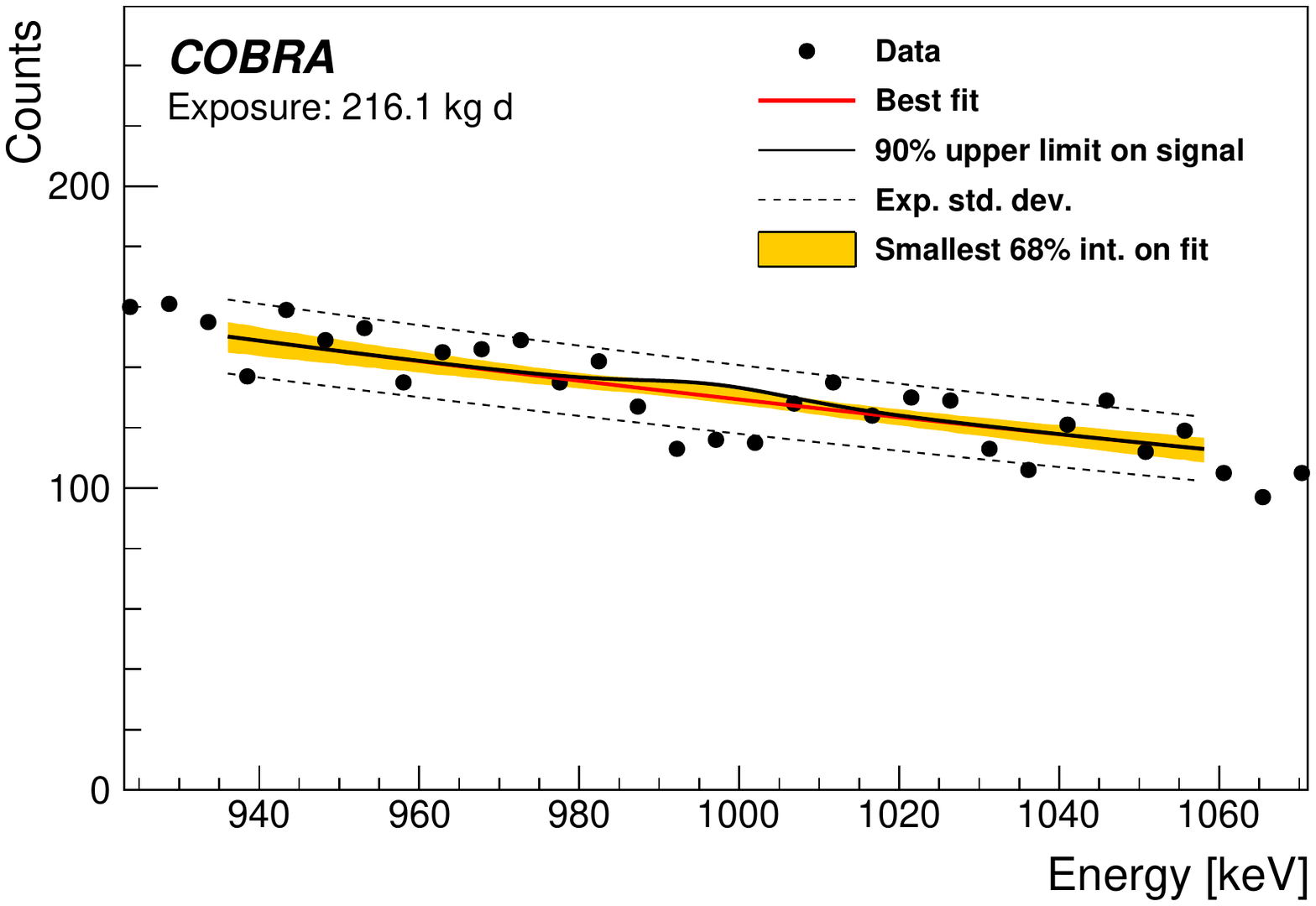}
        \caption{\isotope[70][]{Zn}}
        \label{fig:zn70fit}
   \end{subfigure}
   \begin{subfigure}[]{0.49\textwidth}
        \includegraphics[width=\textwidth]{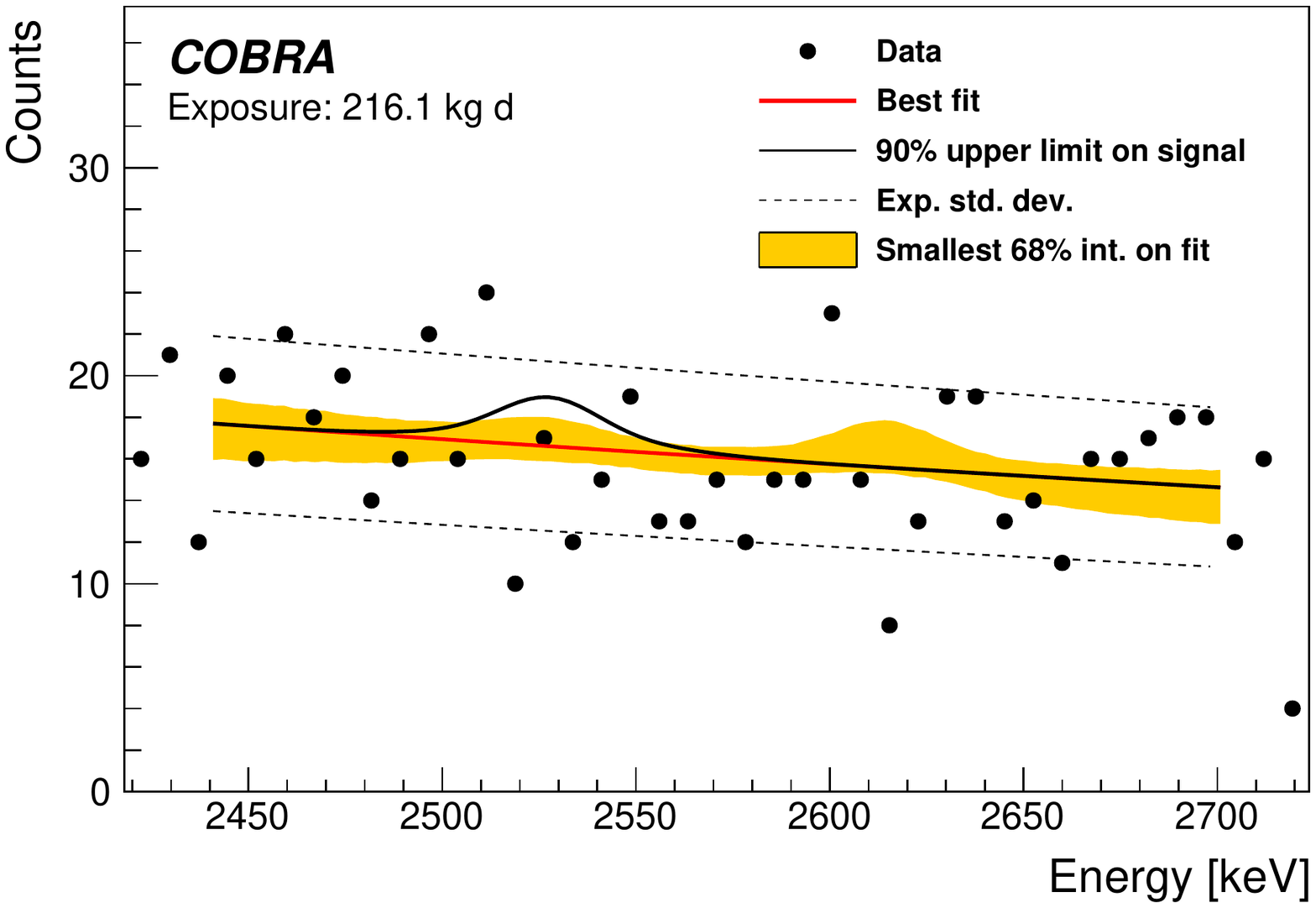}
        \caption{\isotope[130][]{Te}, including the \isotope[208][]{Tl} line}
        \label{fig:te130fit}
   \end{subfigure}
   \begin{subfigure}[]{0.49\textwidth}
        \includegraphics[width=\textwidth]{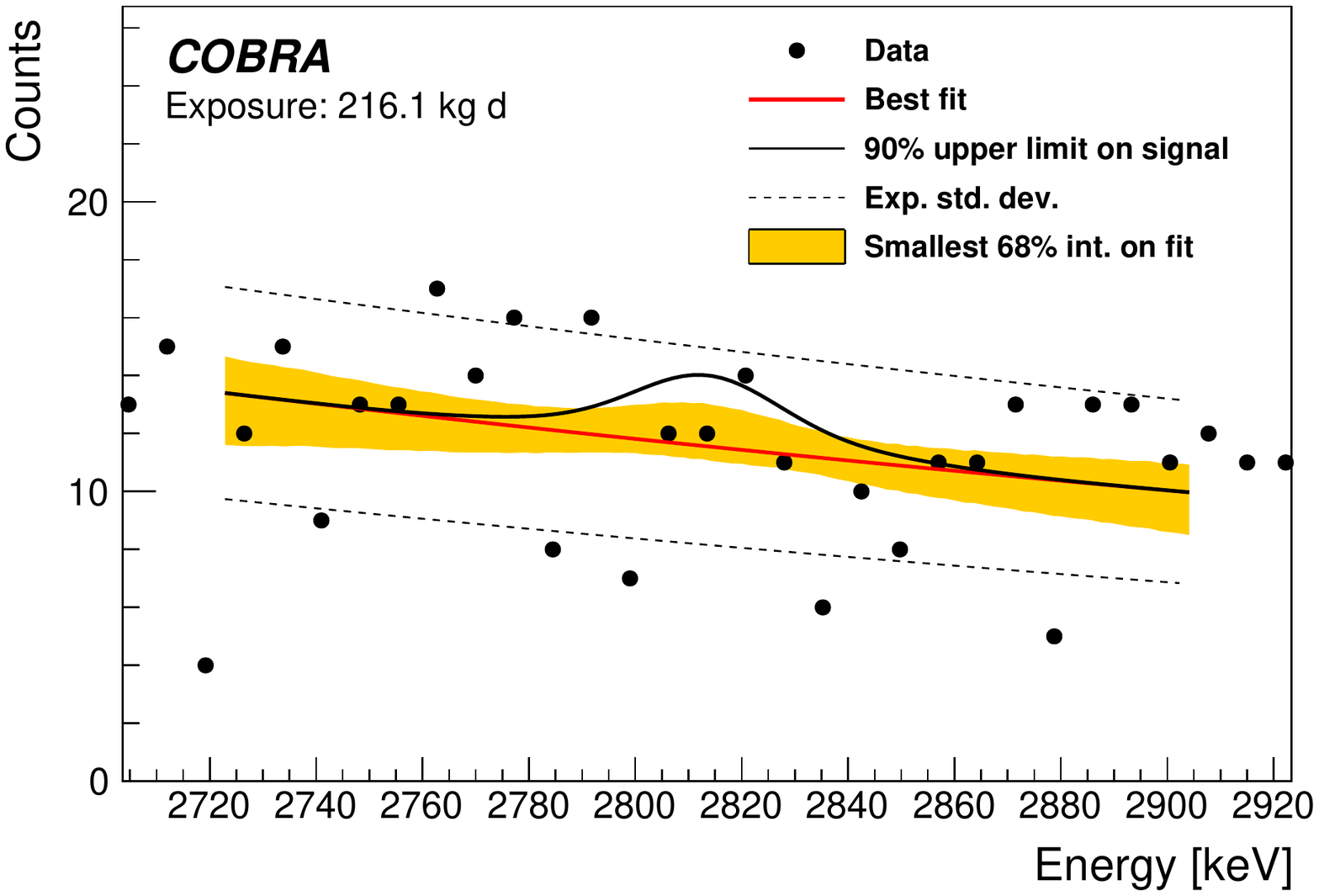}
        \caption{\isotope[116][]{Cd}}
        \label{fig:cd116fit}
   \end{subfigure}
   \caption{Energy spectra of all isotopes. Also shown are the best fit results (red) and the corresponding smallest $68$\,\% uncertainty bands of the fits (yellow). The black lines corresponding to signal contributions equal to the $90\,\%$ upper limits. The black line dashed lines indicate the expected statistical fluctuations of the number of counts in each bin.}
\end{figure*}

The energy spectrum is fitted in several ROIs for the isotopes listed in \autoref{tab:priors}. The table lists the exposure after the run selection, the region of interest, the $Q$-values \cite{qval-2012} and the interval of the uniform prior used for each isotope. The energy spectra including the best fit model and the corresponding uncertainty bands are shown in Figs. \ref{fig:cd114fit} to \ref{fig:cd116fit}.
 \begin{table}[htb]\centering 
    \caption{Regions of interests and the corresponding $Q$-values and prior probabilities for each isotope. The priors vary because of different isotopic abundancies and efficiencies}
    \label{tab:priors}
    \begin{tabular}{c
        S[table-format=1.3,table-comparator=true,table-space-text-post={}] 
        S[table-format=1,table-comparator=true,table-space-text-post={}] 
        S[table-format=3,table-comparator=true,table-space-text-post={*}] 
        S[table-format=1e2,table-comparator=true,table-space-text-post={}] 
        }\toprule\toprule
        Isotope\,\, &\multicolumn{0}{c}{$X$}       & \multicolumn{0}{c}{ROI} &\multicolumn{0}{c}{$Q$}  &\multicolumn{0}{c}{ \,$T_{1/2}^{-1}$ prior}\\
                    &\multicolumn{0}{c}{[kg\,d]}                     &\multicolumn{0}{c}{[keV]}      & \multicolumn{0}{c}{[keV]}     &\multicolumn{0}{c}{ [yr$^{-1}$]}\\\toprule
        \isotope[114][]{Cd}& 212.8 & \numrange{483}{584}  &543&\numrange{0}{ 2e-19} \\
        \isotope[128][]{Te}& 216.1&  \numrange{809} {926}&867&\numrange{0}{ 2e-19 }\\
        \isotope[70][]{Zn} & 216.1& \numrange{940}{1062} &997&\numrange{0}{ 1e-16}\\
        \isotope[130][]{Te}& 216.1& \numrange{2440}{2671} &2528&\numrange{0}{ 4e-20}\\
        \isotope[116][]{Cd}& 216.1& \numrange{2723}{2904} &2814&\numrange{0}{ 2e-19} \\\bottomrule\bottomrule
    \end{tabular}
\end{table}

For each ROI, a signal model, $H_1$, and a background only model, $H_0$, are defined. The Bayes factor is defined as
 \begin{equation}
     K=\frac{P(D|H_1)}{P(D|H_0)},
 \end{equation}
where $P(D|H_1)$ is the probability of the data given the signal model and $P(D|H_0)$ the probability of the data given the $H_{0}$ model.
Bayes factors below 1 indicates that the signal hypothesis is disfavored against the background-only hypothesis.

All Bayes factors are well below 1, and thus the signal hypothesis is disfavored for each isotope. The absolute values of $K$ are listed in \autoref{tab:results}.  The observed events in the ROIs range from $288$ (\isotope[116][]{Cd}) to $\numprint{18670}$ (\isotope[114][]{Cd}). The corresponding background indices range from $2.7\,$counts/(keV\,kg\,yr) (\isotope[116][]{Cd}) up to $213.9\,$counts/(keV\,kg\,yr) (\isotope[114][]{Cd}). 

For \isotope[130][]{Te} the \isotope[208][]{Tl} line at $\numprint{2614}\,$keV is included in the analysis to scan for possible contributions of the thorium decay chain. 
\begin{table}[t!]
\caption{Results of the Bayesian signal estimation after folding all systematic uncertainties. The second column shows the natural abundance of atoms per kg which is used for the signal estimation, for its uncertainty see \autoref{tab:systematic}. In the third column the background index $b$ for the different ROIs is presented. The fourth column reports the limit at 90\,\% credibility and the last column reports the calculated Bayes factor.}
\label{tab:results}
\begin{tabular}{
    c
     S[table-format=1.1,table-comparator=true,table-space-text-post={}] 
     r
     S[table-format=1.1e2,table-comparator=true,table-space-text-post={}]
    c
    }\toprule\toprule
    Isotope             &   \multicolumn{0}{c}{$N$/$10^{23}$}      &  \multicolumn{0}{c}{ $b$ }&\multicolumn{0}{c}{$T_{1/2}\,90\%$ C.L.}& K\\ 
                        &   \multicolumn{0}{c}{$[\frac{\text{atoms}}{\text{kg}}]$}                                       & \multicolumn{0}{r}{ [$\frac{\text{counts}}{\text{keV\,kg\,yr}}$]}& \multicolumn{0}{c}{[$10^{21}\text{yr}$]}& \\\toprule
    \isotope[114][]{Cd} & 6.59                                                    & $ 213.9^{+1}_{-1.7}$  & 1.6& 0.07\\ 
    \isotope[128][]{Te} & 8.08                                                    & $65.5^{+0.5}_{-1.6}$ & 1.9&0.17\\ 
    \isotope[70][]{Zn}  & 0.015                                                   & $45.1^{+0.6}_{-1}$ & 6.8e-3&0.06\\ 
    \isotope[130][]{Te} & 8.62                                                    & $3.6^{+0.1}_{-0.3} $ & 6.1&0.14\\ 
    \isotope[116][]{Cd} & 1.73                                                    & $ 2.7^{+0.1}_{-0.2}$ & 1.1&0.27\\\bottomrule\bottomrule
\end{tabular}
\end{table}
If the signal hypothesis is rejected, a lower limit on the half life at 90\,\% credibility is calculated. For \isotope[114][]{Cd}, the analysis results in a lower limit of $T_{1/2}>1.6\times10^{21}\,$yr. The included annihilation line has a signal strength of $1770\,$counts. For \isotope[128][]{Te} the analysis yields zero events resulting in a limit of $T_{1/2}>1.9\times10^{21}\,$yr. The fit of \isotope[70][]{Zn} prefers zero events which results in a limit of $T_{1/2}>6.8\times10^{18}\,$yr. The \isotope[208][]{Tl} line included in the \isotope[130][]{Te} fit yields zero events, so there is no evidence for any background contribution from the Thorium decay chain. The limit for \isotope[130][]{Te} is $T_{1/2}>6.1\times10^{21}\,$yr. In the \isotope[116][]{Cd} ROI the fit yields zero events for the \Onbb\, mode. The resulting limit is $T_{1/2}>1.1\times10^{21}\,$yr. Switching to a prior uniform in the effective Majorana neutrino mass increases the limits on the half-life by 30--40\%, depending on the isotope.

\autoref{tab:results} reports the abundance per kg detector mass of the analyzed isotope, the total efficiency, the background level in each ROI and the 90\,\% credibility lower limit on the half-life for each isotope.

\section{Conclusion}
A search for \Onbb\, with the COBRA demonstrator setup has been performed using data corresponding to an exposure of $234.7$\,kg\,d. No indication for \Onbb\, have been found and 90\,\% credibility limits have been set for different isotopes. The COBRA demonstrator continues to collect data. A layer of large segmented CdZnTe detectors will be installed in the near future, which is expected to increase the sensitivity considerably. 

\begin{acknowledgments}
    We thank the LNGS for the continuous support of the COBRA experiment, and the AMANDA collaboration, especially R. Wischnewski, for providing us with their FADCs. We acknowledge the  CTU Prague and the ECAP, Erlangen, for valuable discussions. COBRA is supported by the German Research Foundation DFG (Zu123/3 and GO1133/1). 
\end{acknowledgments}

\bibliography{cobra_lateral_surface_event_discrimination}

\begin{thebibliography}{24}%
\makeatletter
\providecommand \@ifxundefined [1]{%
 \@ifx{#1\undefined}
}%
\providecommand \@ifnum [1]{%
 \ifnum #1\expandafter \@firstoftwo
 \else \expandafter \@secondoftwo
 \fi
}%
\providecommand \@ifx [1]{%
 \ifx #1\expandafter \@firstoftwo
 \else \expandafter \@secondoftwo
 \fi
}%
\providecommand \natexlab [1]{#1}%
\providecommand \enquote  [1]{``#1''}%
\providecommand \bibnamefont  [1]{#1}%
\providecommand \bibfnamefont [1]{#1}%
\providecommand \citenamefont [1]{#1}%
\providecommand \href@noop [0]{\@secondoftwo}%
\providecommand \href [0]{\begingroup \@sanitize@url \@href}%
\providecommand \@href[1]{\@@startlink{#1}\@@href}%
\providecommand \@@href[1]{\endgroup#1\@@endlink}%
\providecommand \@sanitize@url [0]{\catcode `\\12\catcode `\$12\catcode
  `\&12\catcode `\#12\catcode `\^12\catcode `\_12\catcode `\%12\relax}%
\providecommand \@@startlink[1]{}%
\providecommand \@@endlink[0]{}%
\providecommand \url  [0]{\begingroup\@sanitize@url \@url }%
\providecommand \@url [1]{\endgroup\@href {#1}{\urlprefix }}%
\providecommand \urlprefix  [0]{URL }%
\providecommand \Eprint [0]{\href }%
\providecommand \doibase [0]{http://dx.doi.org/}%
\providecommand \selectlanguage [0]{\@gobble}%
\providecommand \bibinfo  [0]{\@secondoftwo}%
\providecommand \bibfield  [0]{\@secondoftwo}%
\providecommand \translation [1]{[#1]}%
\providecommand \BibitemOpen [0]{}%
\providecommand \bibitemStop [0]{}%
\providecommand \bibitemNoStop [0]{.\EOS\space}%
\providecommand \EOS [0]{\spacefactor3000\relax}%
\providecommand \BibitemShut  [1]{\csname bibitem#1\endcsname}%
\let\auto@bib@innerbib\@empty
\bibitem [{\citenamefont {Goeppert-Mayer}(1935)}]{PhysRev.48.512}%
  \BibitemOpen
  \bibfield  {author} {\bibinfo {author} {\bibfnamefont {M.}~\bibnamefont
  {Goeppert-Mayer}},\ }\href {\doibase 10.1103/PhysRev.48.512} {\bibfield
  {journal} {\bibinfo  {journal} {Phys. Rev.}\ }\textbf {\bibinfo {volume}
  {48}},\ \bibinfo {pages} {512} (\bibinfo {year} {1935})}\BibitemShut
  {NoStop}%
\bibitem [{\citenamefont {Barabash}\ and\ \citenamefont
  {Brudanin}(2011)}]{Barabash:2010bd}%
  \BibitemOpen
  \bibfield  {author} {\bibinfo {author} {\bibfnamefont {A.~S.}\ \bibnamefont
  {Barabash}}\ and\ \bibinfo {author} {\bibfnamefont {V.~B.}\ \bibnamefont
  {Brudanin}} (\bibinfo {collaboration} {NEMO}),\ }\href {\doibase
  10.1134/S1063778811020062} {\bibfield  {journal} {\bibinfo  {journal} {Phys.
  Atom. Nucl.}\ }\textbf {\bibinfo {volume} {74}},\ \bibinfo {pages} {312}
  (\bibinfo {year} {2011})},\ \Eprint {http://arxiv.org/abs/1002.2862}
  {arXiv:1002.2862 [nucl-ex]} \BibitemShut {NoStop}%
\bibitem [{\citenamefont {Barabash}(2010)}]{PhysRevC.81.035501}%
  \BibitemOpen
  \bibfield  {author} {\bibinfo {author} {\bibfnamefont {A.~S.}\ \bibnamefont
  {Barabash}},\ }\href {\doibase 10.1103/PhysRevC.81.035501} {\bibfield
  {journal} {\bibinfo  {journal} {Phys. Rev. C}\ }\textbf {\bibinfo {volume}
  {81}},\ \bibinfo {pages} {035501} (\bibinfo {year} {2010})}\BibitemShut
  {NoStop}%
\bibitem [{\citenamefont {Thomas}\ \emph {et~al.}(2008)\citenamefont {Thomas},
  \citenamefont {Pattrick}, \citenamefont {Crowther}, \citenamefont
  {Blagburn},\ and\ \citenamefont {Gilmour}}]{PhysRevC.78.054606}%
  \BibitemOpen
  \bibfield  {author} {\bibinfo {author} {\bibfnamefont {H.~V.}\ \bibnamefont
  {Thomas}}, \bibinfo {author} {\bibfnamefont {R.~A.~D.}\ \bibnamefont
  {Pattrick}}, \bibinfo {author} {\bibfnamefont {S.~A.}\ \bibnamefont
  {Crowther}}, \bibinfo {author} {\bibfnamefont {D.~J.}\ \bibnamefont
  {Blagburn}}, \ and\ \bibinfo {author} {\bibfnamefont {J.~D.}\ \bibnamefont
  {Gilmour}},\ }\href {\doibase 10.1103/PhysRevC.78.054606} {\bibfield
  {journal} {\bibinfo  {journal} {Phys. Rev. C}\ }\textbf {\bibinfo {volume}
  {78}},\ \bibinfo {pages} {054606} (\bibinfo {year} {2008})}\BibitemShut
  {NoStop}%
\bibitem [{\citenamefont {Furry}(1939)}]{PhysRev.56.1184}%
  \BibitemOpen
  \bibfield  {author} {\bibinfo {author} {\bibfnamefont {W.~H.}\ \bibnamefont
  {Furry}},\ }\href {\doibase 10.1103/PhysRev.56.1184} {\bibfield  {journal}
  {\bibinfo  {journal} {Phys. Rev.}\ }\textbf {\bibinfo {volume} {56}},\
  \bibinfo {pages} {1184} (\bibinfo {year} {1939})}\BibitemShut {NoStop}%
\bibitem [{\citenamefont {Majorana}(1937)}]{Majorana:1937vz}%
  \BibitemOpen
  \bibfield  {author} {\bibinfo {author} {\bibfnamefont {E.}~\bibnamefont
  {Majorana}},\ }\href {\doibase 10.1007/BF02961314} {\bibfield  {journal}
  {\bibinfo  {journal} {Nuovo Cim.}\ }\textbf {\bibinfo {volume} {14}},\
  \bibinfo {pages} {171} (\bibinfo {year} {1937})}\BibitemShut {NoStop}%
\bibitem [{\citenamefont {Agostini}\ \emph {et~al.}(2013)\citenamefont
  {Agostini} \emph {et~al.}}]{Agostini:2013mzu}%
  \BibitemOpen
  \bibfield  {author} {\bibinfo {author} {\bibfnamefont {M.}~\bibnamefont
  {Agostini}} \emph {et~al.} (\bibinfo {collaboration} {GERDA}),\ }\href
  {\doibase 10.1103/PhysRevLett.111.122503} {\bibfield  {journal} {\bibinfo
  {journal} {Phys.Rev.Lett.}\ }\textbf {\bibinfo {volume} {111}},\ \bibinfo
  {pages} {122503} (\bibinfo {year} {2013})},\ \Eprint
  {http://arxiv.org/abs/1307.4720} {arXiv:1307.4720 [nucl-ex]} \BibitemShut
  {NoStop}%
\bibitem [{\citenamefont {Asakura}\ \emph {et~al.}(2015)\citenamefont {Asakura}
  \emph {et~al.}}]{kamland_zen_2014}%
  \BibitemOpen
  \bibfield  {author} {\bibinfo {author} {\bibfnamefont {K.}~\bibnamefont
  {Asakura}} \emph {et~al.} (\bibinfo {collaboration} {KamLAND-Zen}),\
  }\bibfield  {booktitle} {\emph {\bibinfo {booktitle} {{Proceedings, 26th
  International Conference on Neutrino Physics and Astrophysics (Neutrino
  2014)}}},\ }\href {\doibase 10.1063/1.4915593} {\bibfield  {journal}
  {\bibinfo  {journal} {AIP Conf. Proc.}\ }\textbf {\bibinfo {volume} {1666}},\
  \bibinfo {pages} {170003} (\bibinfo {year} {2015})},\ \Eprint
  {http://arxiv.org/abs/1409.0077} {arXiv:1409.0077 [physics.ins-det]}
  \BibitemShut {NoStop}%
\bibitem [{\citenamefont {Albert}\ \emph {et~al.}(2014)\citenamefont {Albert}
  \emph {et~al.}}]{exo_200_results_2014}%
  \BibitemOpen
  \bibfield  {author} {\bibinfo {author} {\bibfnamefont {J.~B.}\ \bibnamefont
  {Albert}} \emph {et~al.} (\bibinfo {collaboration} {EXO-200}),\ }\href
  {\doibase 10.1038/nature13432} {\bibfield  {journal} {\bibinfo  {journal}
  {Nature}\ }\textbf {\bibinfo {volume} {510}},\ \bibinfo {pages} {229}
  (\bibinfo {year} {2014})},\ \Eprint {http://arxiv.org/abs/1402.6956}
  {arXiv:1402.6956 [nucl-ex]} \BibitemShut {NoStop}%
\bibitem [{\citenamefont {Alfonso}\ \emph {et~al.}(2015)\citenamefont {Alfonso}
  \emph {et~al.}}]{Alfonso:2015wka}%
  \BibitemOpen
  \bibfield  {author} {\bibinfo {author} {\bibfnamefont {K.}~\bibnamefont
  {Alfonso}} \emph {et~al.} (\bibinfo {collaboration} {CUORE Collaboration}),\
  }\href {\doibase 10.1103/PhysRevLett.115.102502} {\bibfield  {journal}
  {\bibinfo  {journal} {Phys. Rev. Lett.}\ }\textbf {\bibinfo {volume} {115}},\
  \bibinfo {pages} {102502} (\bibinfo {year} {2015})}\BibitemShut {NoStop}%
\bibitem [{\citenamefont {Danevich}\ \emph {et~al.}(2003)\citenamefont
  {Danevich} \emph {et~al.}}]{PhysRevC.68.035501}%
  \BibitemOpen
  \bibfield  {author} {\bibinfo {author} {\bibfnamefont {F.~A.}\ \bibnamefont
  {Danevich}} \emph {et~al.},\ }\href {\doibase 10.1103/PhysRevC.68.035501}
  {\bibfield  {journal} {\bibinfo  {journal} {Phys. Rev. C}\ }\textbf {\bibinfo
  {volume} {68}},\ \bibinfo {pages} {035501} (\bibinfo {year}
  {2003})}\BibitemShut {NoStop}%
\bibitem [{\citenamefont {Belli}\ \emph {et~al.}(2008)\citenamefont {Belli}
  \emph {et~al.}}]{cdwo}%
  \BibitemOpen
  \bibfield  {author} {\bibinfo {author} {\bibfnamefont {P.}~\bibnamefont
  {Belli}} \emph {et~al.},\ }\href {\doibase 10.1140/epja/i2008-10593-6}
  {\bibfield  {journal} {\bibinfo  {journal} {The European Physical Journal A}\
  }\textbf {\bibinfo {volume} {36}},\ \bibinfo {pages} {167} (\bibinfo {year}
  {2008})}\BibitemShut {NoStop}%
\bibitem [{\citenamefont {Belli}\ \emph {et~al.}(2011)\citenamefont {Belli}
  \emph {et~al.}}]{0954-3899-38-11-115107}%
  \BibitemOpen
  \bibfield  {author} {\bibinfo {author} {\bibfnamefont {P.}~\bibnamefont
  {Belli}} \emph {et~al.},\ }\href
  {http://stacks.iop.org/0954-3899/38/i=11/a=115107} {\bibfield  {journal}
  {\bibinfo  {journal} {Journal of Physics G: Nuclear and Particle Physics}\
  }\textbf {\bibinfo {volume} {38}},\ \bibinfo {pages} {115107} (\bibinfo
  {year} {2011})}\BibitemShut {NoStop}%
\bibitem [{\citenamefont {Maneschg}(2015)}]{Maneschg2015188}%
  \BibitemOpen
  \bibfield  {author} {\bibinfo {author} {\bibfnamefont {W.}~\bibnamefont
  {Maneschg}},\ }\href {\doibase 10.1016/j.nuclphysbps.2015.02.039} {\bibfield
  {journal} {\bibinfo  {journal} {Nuclear and Particle Physics Proceedings}\
  }\textbf {\bibinfo {volume} {260}},\ \bibinfo {pages} {188 } (\bibinfo {year}
  {2015})},\ \bibinfo {note} {the 13th International Workshop on Tau Lepton
  Physics (Tau2014)}\BibitemShut {NoStop}%
\bibitem [{\citenamefont {Ebert}\ \emph
  {et~al.}(2016{\natexlab{a}})\citenamefont {Ebert} \emph
  {et~al.}}]{Ebert:2015uta}%
  \BibitemOpen
  \bibfield  {author} {\bibinfo {author} {\bibfnamefont {J.}~\bibnamefont
  {Ebert}} \emph {et~al.},\ }\href {\doibase
  http://dx.doi.org/10.1016/j.nima.2015.10.079} {\bibfield  {journal} {\bibinfo
   {journal} {Nuclear Instruments and Methods in Physics Research Section A:
  Accelerators, Spectrometers, Detectors and Associated Equipment}\ }\textbf
  {\bibinfo {volume} {807}},\ \bibinfo {pages} {114 } (\bibinfo {year}
  {2016}{\natexlab{a}})}\BibitemShut {NoStop}%
\bibitem [{\citenamefont {Ebert}\ \emph
  {et~al.}(2016{\natexlab{b}})\citenamefont {Ebert} \emph
  {et~al.}}]{Ebert:2015aia}%
  \BibitemOpen
  \bibfield  {author} {\bibinfo {author} {\bibfnamefont {J.}~\bibnamefont
  {Ebert}} \emph {et~al.} (\bibinfo {collaboration} {COBRA}),\ }\href {\doibase
  10.1016/j.nima.2016.03.012} {\bibfield  {journal} {\bibinfo  {journal} {Nucl.
  Instrum. Meth.}\ }\textbf {\bibinfo {volume} {A821}},\ \bibinfo {pages} {109}
  (\bibinfo {year} {2016}{\natexlab{b}})},\ \Eprint
  {http://arxiv.org/abs/1508.03217} {arXiv:1508.03217 [physics.ins-det]}
  \BibitemShut {NoStop}%
\bibitem [{\citenamefont {Luke}(1994)}]{cpg94}%
  \BibitemOpen
  \bibfield  {author} {\bibinfo {author} {\bibfnamefont {P.~N.}\ \bibnamefont
  {Luke}},\ }\href {\doibase 10.1063/1.112523} {\bibfield  {journal} {\bibinfo
  {journal} {Applied Physics Letters}\ }\textbf {\bibinfo {volume} {65}},\
  \bibinfo {pages} {2884} (\bibinfo {year} {1994})}\BibitemShut {NoStop}%
\bibitem [{\citenamefont {Fritts}\ \emph {et~al.}(2013)\citenamefont {Fritts}
  \emph {et~al.}}]{FrittsCPG}%
  \BibitemOpen
  \bibfield  {author} {\bibinfo {author} {\bibfnamefont {M.}~\bibnamefont
  {Fritts}} \emph {et~al.},\ }\href {\doibase
  http://dx.doi.org/10.1016/j.nima.2013.01.004} {\bibfield  {journal} {\bibinfo
   {journal} {Nucl. Instrum. Meth. A}\ }\textbf {\bibinfo {volume} {708}},\
  \bibinfo {pages} {1 } (\bibinfo {year} {2013})}\BibitemShut {NoStop}%
\bibitem [{\citenamefont {Schulz}(2011)}]{schulz}%
  \BibitemOpen
  \bibfield  {author} {\bibinfo {author} {\bibfnamefont {O.}~\bibnamefont
  {Schulz}},\ }\emph {\bibinfo {title} {Exploration of new Data Acquisition and
  Background Reduction Techniques for the COBRA Experiment}},\ \href
  {http://hdl.handle.net/2003/29108} {Ph.D. thesis},\ \bibinfo  {school}
  {Technische Universit{\"a}t Dortmund} (\bibinfo {year} {2011}),\ \bibinfo
  {note} {{\tt http://hdl.handle.net/2003/29108}}\BibitemShut {NoStop}%
\bibitem [{\citenamefont {Fritts}\ \emph {et~al.}(2014)\citenamefont {Fritts}
  \emph {et~al.}}]{fritts_tebr}%
  \BibitemOpen
  \bibfield  {author} {\bibinfo {author} {\bibfnamefont {M.}~\bibnamefont
  {Fritts}} \emph {et~al.} (\bibinfo {collaboration} {COBRA}),\ }\href
  {\doibase 10.1016/j.nima.2014.02.038} {\bibfield  {journal} {\bibinfo
  {journal} {Nucl. Instrum. Meth.}\ }\textbf {\bibinfo {volume} {A749}},\
  \bibinfo {pages} {27} (\bibinfo {year} {2014})},\ \Eprint
  {http://arxiv.org/abs/1401.5844} {arXiv:1401.5844 [nucl-ex]} \BibitemShut
  {NoStop}%
\bibitem [{\citenamefont {{Caldwell}}\ \emph {et~al.}(2009)\citenamefont
  {{Caldwell}}, \citenamefont {{Koll{\'a}r}},\ and\ \citenamefont
  {{Kr{\"o}ninger}}}]{BAT}%
  \BibitemOpen
  \bibfield  {author} {\bibinfo {author} {\bibfnamefont {A.}~\bibnamefont
  {{Caldwell}}}, \bibinfo {author} {\bibfnamefont {D.}~\bibnamefont
  {{Koll{\'a}r}}}, \ and\ \bibinfo {author} {\bibfnamefont {K.}~\bibnamefont
  {{Kr{\"o}ninger}}},\ }\href {\doibase 10.1016/j.cpc.2009.06.026} {\bibfield
  {journal} {\bibinfo  {journal} {Comp. Phys. Commu.}\ }\textbf {\bibinfo
  {volume} {180}},\ \bibinfo {pages} {2197} (\bibinfo {year} {2009})},\ \Eprint
  {http://arxiv.org/abs/0808.2552} {arXiv:0808.2552 [physics.data-an]}
  \BibitemShut {NoStop}%
\bibitem [{\citenamefont {Wang}\ \emph {et~al.}(2012)\citenamefont {Wang} \emph
  {et~al.}}]{qval-2012}%
  \BibitemOpen
  \bibfield  {author} {\bibinfo {author} {\bibfnamefont {M.}~\bibnamefont
  {Wang}} \emph {et~al.},\ }\href
  {http://stacks.iop.org/1674-1137/36/i=12/a=003} {\bibfield  {journal}
  {\bibinfo  {journal} {Chinese Physics C}\ }\textbf {\bibinfo {volume} {36}},\
  \bibinfo {pages} {1603} (\bibinfo {year} {2012})}\BibitemShut {NoStop}%
\bibitem [{\citenamefont {Allison}\ \emph {et~al.}(2006)\citenamefont {Allison}
  \emph {et~al.}}]{geant}%
  \BibitemOpen
  \bibfield  {author} {\bibinfo {author} {\bibfnamefont {J.}~\bibnamefont
  {Allison}} \emph {et~al.},\ }\href {\doibase 10.1109/TNS.2006.869826}
  {\bibfield  {journal} {\bibinfo  {journal} {Nuclear Science, IEEE
  Transactions on}\ }\textbf {\bibinfo {volume} {53}},\ \bibinfo {pages} {270}
  (\bibinfo {year} {2006})}\BibitemShut {NoStop}%
\bibitem [{\citenamefont {Ponkratenko}\ \emph {et~al.}(2001)\citenamefont
  {Ponkratenko}, \citenamefont {Tretyak},\ and\ \citenamefont
  {Zdesenko}}]{decay0}%
  \BibitemOpen
  \bibfield  {author} {\bibinfo {author} {\bibfnamefont {O.~A.}\ \bibnamefont
  {Ponkratenko}}, \bibinfo {author} {\bibfnamefont {V.~I.}\ \bibnamefont
  {Tretyak}}, \ and\ \bibinfo {author} {\bibfnamefont {Y.~G.}\ \bibnamefont
  {Zdesenko}},\ }\href {\doibase 10.1134/1.855784} {\bibfield  {journal}
  {\bibinfo  {journal} {Physics of Atomic Nuclei}\ }\textbf {\bibinfo {volume}
  {63}},\ \bibinfo {pages} {1282} (\bibinfo {year} {2001})}\BibitemShut
  {NoStop}%
\end{thebibliography}%
\bibliographystyle{apsrev4-1}

\end{document}